\documentclass[pre,showpacs,twocolumn]{revtex4-1}
\usepackage{latexsym}
\usepackage{amsfonts}
\usepackage{amsmath}
\usepackage{multirow}
\usepackage{graphicx}
\usepackage{color}

\def\E{{\mathbb E}}
\def\Var{{\mathbb{V}\mathrm{ar}}}

\def\E{{\mathbb E}}

\begin{document}
\title{Optimal parameters for anomalous diffusion exponent estimation from noisy data}

\author{Yann Lanoiselee}
\email{yann.lanoiselee@polytechnique.edu}
\author{Denis S. Grebenkov}
\email{denis.grebenkov@polytechnique.edu}
\affiliation{
Laboratoire de Physique de la Mati\`{e}re Condens\'{e}e (UMR 7643), \\ 
CNRS -- Ecole Polytechnique, 91128 Palaiseau, France}

\author{Grzegorz Sikora}
\email{grzegorz.sikora@pwr.edu.pl}
\author{Aleksandra Grzesiek}
\email{aleksandra.grzesiek@pwr.edu.pl}
\author{Agnieszka Wy{\l}oma{\'n}ska}
\email{agnieszka.wylomanska@pwr.edu.pl}
\affiliation{Faculty of Pure and Applied Mathematics, Hugo Steinhaus Center,\\
Wroclaw University of Science and Technology, Wyb. Wyspianskiego 27, 50-370 Wroclaw, Poland 
}

\begin{abstract}
The most common way of estimating the anomalous diffusion exponent from single-particle trajectories consists in a linear fitting of the dependence of the time averaged mean square displacement on the lag time at the log-log scale. However, various measurement noises that are unavoidably present in experimental data, can strongly deteriorate the quality of this estimation procedure and bias the estimated exponent. To investigate the impact of noises and to improve the estimation quality, we compare three approaches for estimating the anomalous diffusion exponent and check their efficiency on fractional Brownian motion corrupted by Gaussian noise. We discuss how the parameters of this anomalous diffusion model and the parameters of the estimation techniques influence the estimated exponent. We show that the conventional linear fitting is the least optimal method for the analysis of noisy data.
\end{abstract}
\maketitle

\section{Introduction}

Anomalous diffusion processes are widely discussed in the literature,
in particular, in the context of single-particle trajectories analysis
\cite{Arcizet08,Metzler09c,Tejedor2010,Magdziarz2011,greb11,greb12,greb13,Gal2013a,Meroz2013,Turkcan2013,eldad2015,Lanoiselee2016}.
The anomalous diffusive behavior is manifested by non-linear time
growth of the mean square displacement (MSD), $\langle X^2(\tau) \rangle
\simeq 2D_\beta \tau^\beta$, where $\beta$ is the anomalous diffusion
exponent, $D_{\beta}$ is the generalized diffusion coefficient (in
units m$^2$/s$^{\beta}$), and $\langle.\rangle$ denotes the (ensemble)
average over the probability distribution of $X(\tau)$. Depending on
the $\beta$ parameter one can distinguish between sub-diffusive ($\beta
< 1$), diffusive ($\beta = 1$), and super-diffusive ($\beta > 1$)
behavior,
\cite{Tolic04,Golding06,Wilhelm08,Szymanski09,Metzler09c,Sackmann10,Jeon11,Bertseva12,Bressloff13}.
However, due to a limited number of trajectories in many experiments,
the ensemble average (EA) MSD needs to be replaced by the time average
(TA) MSD calculated from a single trajectory.  For a vector of
observations $X(1),X(2),\ldots,X(N)$ of length $N$, the TAMSD at the
lag time $\tau$ is defined as
\begin{equation}
\label{eq:msd}
M_N(\tau)=\frac{1}{N-\tau}\sum_{i=1}^{N-\tau}(X(i+\tau)-X(i))^2.
\end{equation}

For an ergodic process with stationary increments, TAMSD converges to
EAMSD in the limit $N\to\infty$, $M_{N\to\infty}(\tau)=\langle
X^2(\tau)\rangle $, i.e., the distribution of TAMSD converges to a
Dirac delta function centered on the value of EAMSD.  Consequently,
for $1\leq\tau\ll N$ the mean TAMSD scales as
\begin{eqnarray}\label{eq:MSD_basis}
\langle M_N(\tau)\rangle \simeq 2D_\beta\tau^{\beta}.
\end{eqnarray}
The TAMSD is one of the classical tools used for estimation of the
anomalous diffusion exponent $\beta$.  The procedure of estimation is
simple: the TAMSD is plotted versus the lag time $\tau$ at the log-log
scale and the estimated $\beta$ parameter is the slope of the expected
straight line, fitted by using the least squares method
\cite{Gal2013a,eldad2015}. 

The classical pure anomalous diffusion models include fractional
Brownian motion (fBm) \cite{beran,man68}, fractional L\'evy stable
motion \cite{lm1} and continuous-time random walk
\cite{Metzler00,Metzler04}. In this paper, we focus on
the fBm that is a non-Markovian generalization of Brownian motion and
one of the most fundamental models of stochastic motion. Specifically,
it is the only self-similar Gaussian process with stationary
increments. The fBm can also be related to generalized Langevin processes with power law decaying friction kernels, an attractive
framework for many physical systems
\citep{KLM,burwer10,Grebenkov11a,burn_sik}.

One of the main statistical challenges in the experimental data
analysis is the proper model recognition and the precise estimation of
the best model parameters.  In this paper, we focus on the estimation
of the parameters of noisy anomalous diffusion in which ``pure''
(i.e. noiseless) fBm is progressively corrupted by Gaussian white
noise. We propose two alternative approaches for anomalous diffusion
exponent estimation and compare them to the common linear fitting on
simulated data. Moreover, we discuss how the parameters of the
considered model influence the estimation results. The similar problem
was discussed in \cite{Mic10,Mic12,ErnKoh} in case of ordinary
Brownian motion.

The rest of the paper is organized as follows: in the next section we
formulate the problem. In section \ref{sec_2} we propose and compare
three approaches for anomalous parameters estimation. In section
\ref{sec:numerical_results} we check the efficiency of the proposed
estimation techniques on simulated data. The last section concludes
the paper.

\section{Problem formulation}

The classical approach for estimating the parameters $D_{\beta}$ and
$\beta$ from Eq. (\ref{eq:MSD_basis}) for ``pure'' anomalous diffusion
consists in a linear fitting. More precisely, the TAMSD first is
calculated from a vector of positions according to
Eq. (\ref{eq:msd}). Then, taking the logarithm of both sides of the
formula (\ref{eq:MSD_basis}) one can estimate the parameters using the
classical least squares method in linear regression. The details of
this approach are presented for instance in
\cite{testexponent}. Usually, the parameters are estimated by using
integer lag-times $\tau\in[1,\tau_{\max}]$. The accuracy of the
estimation decreases as $\tau_{\max}$ gets larger. In spite of its
numerous applications in practice, the approach has some drawbacks.

Even if the experimental data exhibit a behavior adequate to some
theoretical model of anomalous diffusion, it is always disturbed by
measurement noise \cite{Kepten2015}
\begin{eqnarray}\label{eq:theoretical}
X(\tau)=Z(\tau)+\xi(\tau),
\end{eqnarray}
where $Z(\tau)$ is a ``pure'' anomalous diffusion process with
$D_{\beta}$ and $\beta$ parameters, and $\xi(\tau)$ denotes noise,
which is assumed to be independent from $Z(\tau)$ and normally
distributed with mean zero and variance $\sigma^2$. The EAMSD reads
then
\begin{equation}\label{eq:MSD_plus_noise}
\langle X^2(\tau)\rangle=2D_\beta \tau^{\beta}+\sigma^2.
\end{equation}

\begin{figure}
\begin{center}
\includegraphics[width=0.5\textwidth]{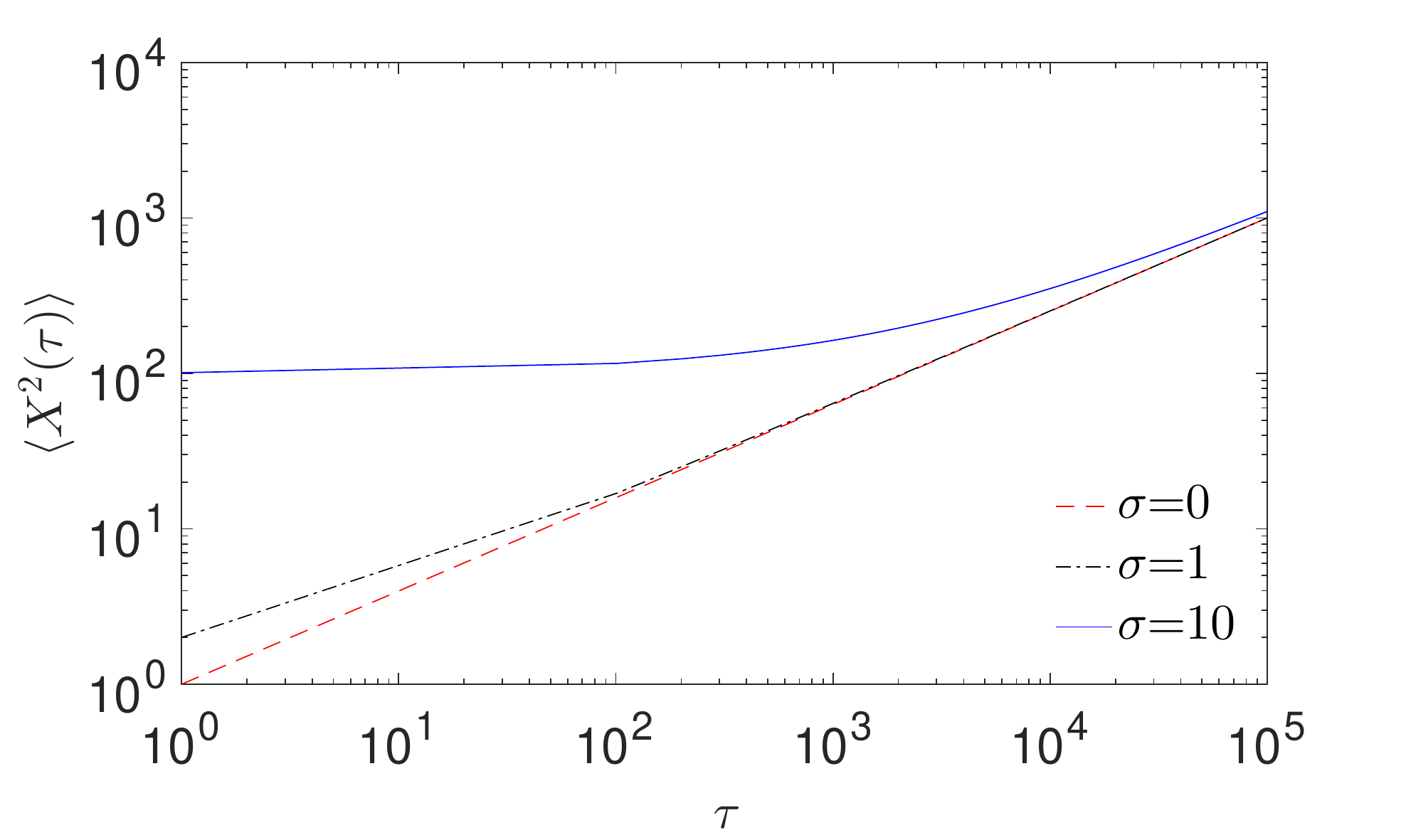}
\end{center}
\caption{
The MSD of the fractional Brownian motion with $\beta=0.6$ and
$D_\beta = 1/2$, corrupted by white noise with three different values
of the standard deviation $\sigma=\lbrace0,1,10\rbrace$.  Arbitrary
units are used.}
\label{fig:plot_MSD_plus_noise}
\end{figure}

Figure \ref{fig:plot_MSD_plus_noise} shows that the noise term
$\sigma^2$ makes the EAMSD (as well as the TAMSD) flat, until the
contribution from anomalous diffusion becomes dominant:
$2D_\beta\tau^{\beta}\gg \sigma^2$. To avoid such a noise dominated
region, it is natural to perform the fitting from $\tau_{\min}$ to
$\tau_{\max}$, with some $\tau_{\min} > 1$.  This is the first problem
discussed in this paper. We check by simulations how the noise term
$\sigma^2$ influences the estimation results and how the selection of
the $\tau_{\min}$ and $\tau_{\max}$ in the classical estimation
algorithm can change the estimation efficiency. 

The second considered problem can be formulated as follows: even if
fitting is performed over the window from $\tau_{\min}$ to
$\tau_{\max}$, it is not enough to get efficient estimators of
$D_{\beta}$ and $\beta$ from the {\it linear} fit, because departures
from the linear shape of MSD is increasing with $\sigma$ (see
Fig. \ref{fig:plot_MSD_plus_noise}). In this paper, we propose two
alternative approaches for estimating $\beta$ and $D_\beta$ via a
non-linear fitting. To our knowledge, the non-linear fitting approach
to estimating anomalous diffusion parameters was not systematically
studied yet.  Both approaches assume the toy model defined in
Eq. (\ref{eq:theoretical}), i.e. the noise term is taken into
consideration. Then, we compare the estimation results for the
proposed methods with the classical method where the model is just
anomalous diffusive process $Z(\tau)$. Moreover, we check also the
influence of $\sigma$, $\tau_{\min},$ and $\tau_{\max}$ on the
estimation results for two approaches. The simulations will be
presented for the selected anomalous diffusion model $Z(\tau)$ in
(\ref{eq:theoretical}), namely fBm, however we would like to highlight
that the problem is relevant for any ergodic process showing anomalous
diffusion.

\section{Anomalous diffusion exponent estimation}
\label{sec_2}

\noindent 
In this section, we describe three approaches used to estimate the
anomalous diffusion exponent $\beta$. Although we focus on the
anomalous diffusion exponent estimation, the presented approaches are
also useful for estimating the diffusion parameter $D_{\beta}$.

\subsection{Approach I}\label{linear_fit}

\noindent 
The classical Approach I consists in taking the logarithm of both
sides of Eq. (\ref{eq:MSD_plus_noise}) and expanding the right-hand
side to the first order of with respect to the small parameter
$\frac{\sigma^2}{2D_\beta \tau^{\beta}} \ll 1$. The relation becomes
\begin{equation}\label{MSD_small_noise}
\ln\left(\langle X^2(\tau)\rangle\right)= \ln(2D_\beta)+ \beta \ln(\tau)+\frac{\sigma^2}{2D_\beta \tau^{\beta}}+O(\tau^{-2\beta}),
\end{equation}
then with the variable $u=\ln(\tau)$ we get
\begin{equation}
\ln\left(\langle X^2(u)\rangle\right)= \ln(2D_\beta)+ \beta u+\frac{\sigma^2}{2D_\beta}e^{-\beta u}+O(e^{-2\beta u}).
\end{equation}
There is a linear dependence of $\ln\left(\langle
X^2(u)\rangle\right)$ on $u$ with an exponentially decaying (in
log-log coordinates) correction related to the noise term
$\sigma^2$. In the limit either of small $\sigma^2$ or large $u$, the
noise effect disappears and the estimation is reduced to a linear
regression.

In this approach, we estimate the $\beta$ exponent in a similar way as
for pure fBm. The details of this approach one can find for instance
in \cite{testexponent} therefore we only sketch the idea. For pure fBm
in order to estimate the anomalous diffusion exponent $\beta$ one
needs to calculate TAMSD from the trajectory $X(1),X(2),\ldots,X(N)$
of length $N$ at the points $\tau_{\min},\ldots,\tau_{\max}$ and
then fit the linear function of a form $\ln({D_{\beta}})+\beta \ln(i)$
to $\ln(M_N(i))$ for $\tau=\tau_{\min},\ldots,\tau_{\max}.$  The exact form of the
estimator $\hat{\beta}$ from the least squares method to linear
fitting reads 
\begin{widetext}
\begin{equation}\label{est_2}
\hat{\beta}=\frac{n\sum_{\tau=\tau_{\min}}^{\tau_{\max}}{\ln(\tau)\ln(M_N(\tau))}-\sum_{\tau=\tau_{\min}}^{\tau_{\max}}{\ln(\tau)}\sum_{\tau=\tau_{\min}}^{\tau_{\max}}{\ln(M_N(\tau))}}{n\sum_{\tau=\tau_{\min}}^{\tau_{\max}}{\ln^2(\tau)}-\left(\sum_{\tau=\tau_{\min}}^{\tau_{\max}}{\ln(\tau)}\right)^2},
\end{equation}
\end{widetext}
where $\Delta\tau=\tau_{\max}-\tau_{\min}$ and $n=\Delta\tau+1.$ In the Approach I, for small values of $\sigma$, we can neglect the
noise term and use the estimator for pure fBm. Other choice of the estimator is possible in the case $\tau_{\min}=1$ \cite{testexponent}. The discussion about the theoretical properties of the modified estimator is presented in Appendix A.

Due to the finite trajectory length $N<\infty$, the selection of
$\tau_{\max}$ too close to $N$ would result in large fluctuations
because of the small number of data points contributing for the
average. Simulations allow checking both the effects of $\tau_{\min}$,
$\tau_{\max}$ on the estimation at a given level of noise (see section
\ref{sec:numerical_results}).

\subsection{Approach II}\label{non-linear_fit}

The Approach II is one of the alternatives to the classical method
presented above. In contrast to Approach I, we estimate the $\beta$
parameter taking into account the noise term $\xi(\tau)$ in the model
(\ref{eq:theoretical}).

The idea is to perform a fitting using the exact formula of the EAMSD
in Eq. (\ref{eq:MSD_plus_noise})
\begin{eqnarray}\label{TAMSD_new}
M_N(\tau)\sim 2D_\beta \tau^{\beta}+\sigma^2, 
\end{eqnarray}
where $\sim$ means the equality in the expected value. Similarly to
Approach I, we calculate $M_N(\tau)$ for the lag times
$\tau_{\min},\ldots,\tau_{\max}$. However, in this case, the fitting
function $\hat{M}(\tau)$ is defined as follows
\begin{equation}\label{f1}
\hat{M}(\tau)=2\hat{D}_\beta\tau^{\hat{\beta}}+\hat{\sigma}^2 ,
\end{equation}
where $\hat{\beta}$, $\hat{D}_\beta$ and $\hat{\sigma}$ are three
fitting parameters.  In order to estimate these parameters, one has to
the minimize the error function
$\Upsilon=\sum_{i=\tau_{\min}}^{\tau_{\max}} (M_N(i)-\hat{M}(i))^2$.
The minimum is found when the gradient of $\Upsilon$ with respect to
the fitting parameters is equal to zero and the error is the lowest.
The function $\hat{M}(\tau)$ has a non-linear dependence on $\tau$
making it dependent on the fitting parameters themselves. In this
case, the explicit expression of an estimator $\hat{\beta}$ is not
accessible and has to be calculated numerically. The non-linear
fitting methodology is described in Appendix \ref{sec:NL_fit}.

\subsection{Approach III}\label{non-linear_removetau_fit}

In the last approach, we make a simple transformation of the TAMSD to
reduce the number of fitting parameters. We take into account the fact
that the fitting is starting at the point $\tau_{\min}$. Thus we
subtract from $M_N(\tau)$, $\tau\in [\tau_{\min},\ldots,
\tau_{\max}]$ the term $M_{N}(\tau_{\min})$ so that the formula
(\ref{TAMSD_new}) reduces to
\begin{equation}\label{msdtau_rem_taumin}
M_N(\tau)-M_N(\tau_{\min})\sim 2{D_{\beta}}\left(\tau^{{\beta}}-\tau_{\min}^{{\beta}}\right).
\end{equation}
This transformation removes the noise term, at the cost of a more
complex dependence on $\beta$. Thus, in the Approach III, we calculate
$M_N(\tau)-M_N(\tau_{\min})$ at lag times
$\tau_{\min},\ldots,\tau_{\max}$ and then fit them by the function
\begin{equation}\label{f2}
\hat{M}(\tau)=2\hat{D}_\beta\left(\tau^{\hat{\beta}}-\tau_{\min}^{\hat{\beta}}\right).
\end{equation}
Similar to Approach II, the function $\hat{M}(\tau)$ is non-linear, so
an iterative procedure is necessary for error minimization.

\section{Optimal parameters for $\beta$ estimation}
\label{sec:numerical_results}

In this section, we discuss the $\beta$ parameter estimation from
Monte Carlo simulations. We simulate single-particle trajectories by
the model in Eq. (\ref{eq:theoretical}), where $Z(\tau)$ is an fBm
with a given $\beta$ and $D_{\beta}=1/2$.  We compare three approaches
and check which one is the most efficient for the $\beta$ parameter
estimation.  Although similar techniques can be used for estimating
the diffusion parameter $D_\beta$, we do not consider this option in
this paper.

We test the three fitting approaches on three representative cases of
the fBm: (i) sub-diffusive anti-persistent motion with $\beta=0.6$,
(ii) diffusive Markovian motion with $\beta=1$ and (iii)
super-diffusive persistent motion with $\beta=1.4$. In order to be
closer to experimental conditions, three levels of noise are tested,
classified from none to high noise with the standard deviation taking
values $\sigma=\lbrace 0,1,10\rbrace$. In every case the fitting is
performed using the TAMSD calculated from a single trajectory. The
performance of each method is measured in terms of the accuracy
$A(\%)$ of the estimation, which is the percentage of the estimated
exponent which falls in the range $\beta-0.2<\hat{\beta}<\beta+0.2$.
The quantity $A$ is calculated from the estimated distribution of
$\hat{\beta}$ obtained from $M=1000$ realizations.

\begin{figure*}
\includegraphics[width=0.3\textwidth]{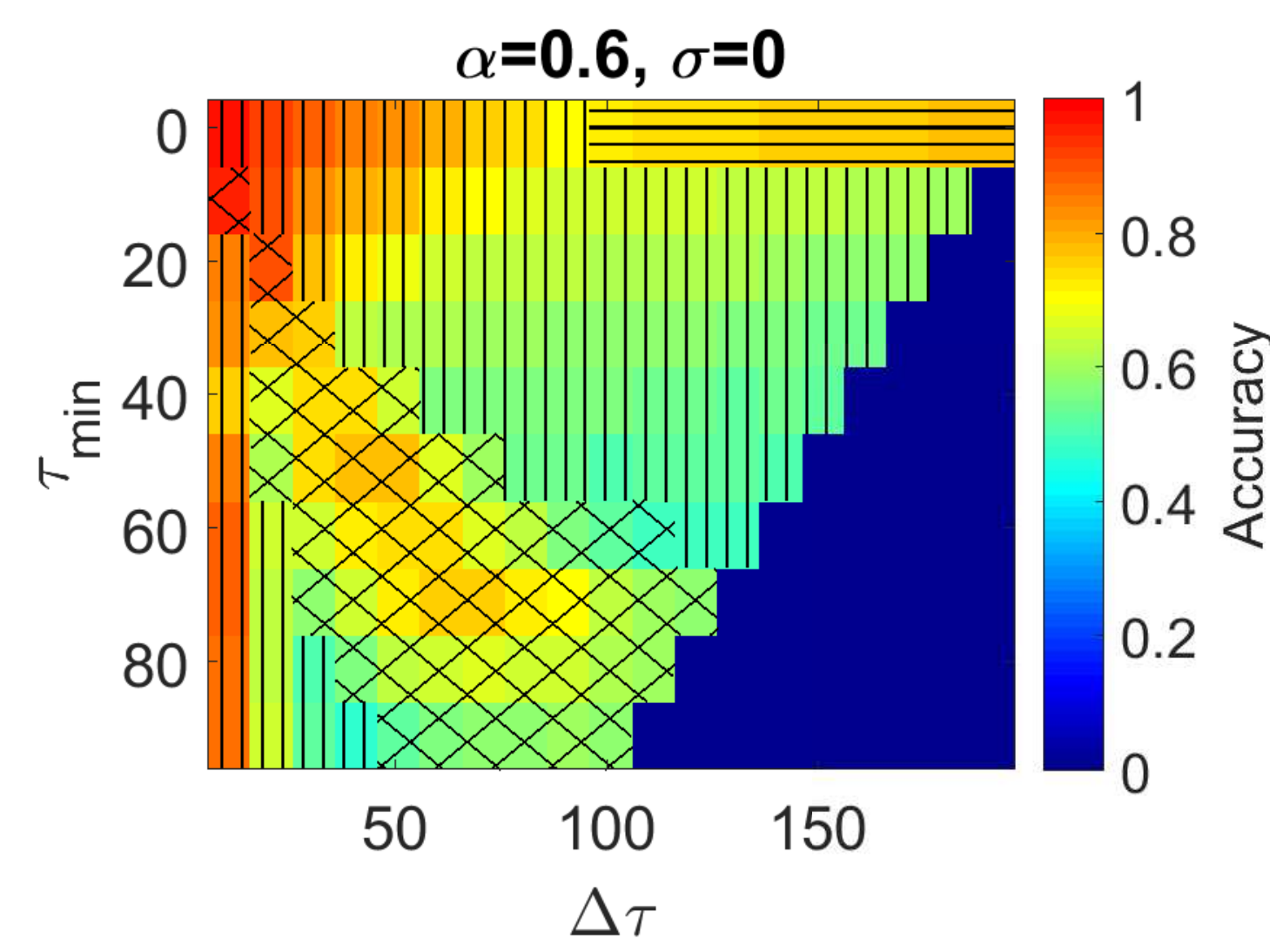}
\includegraphics[width=0.3\textwidth]{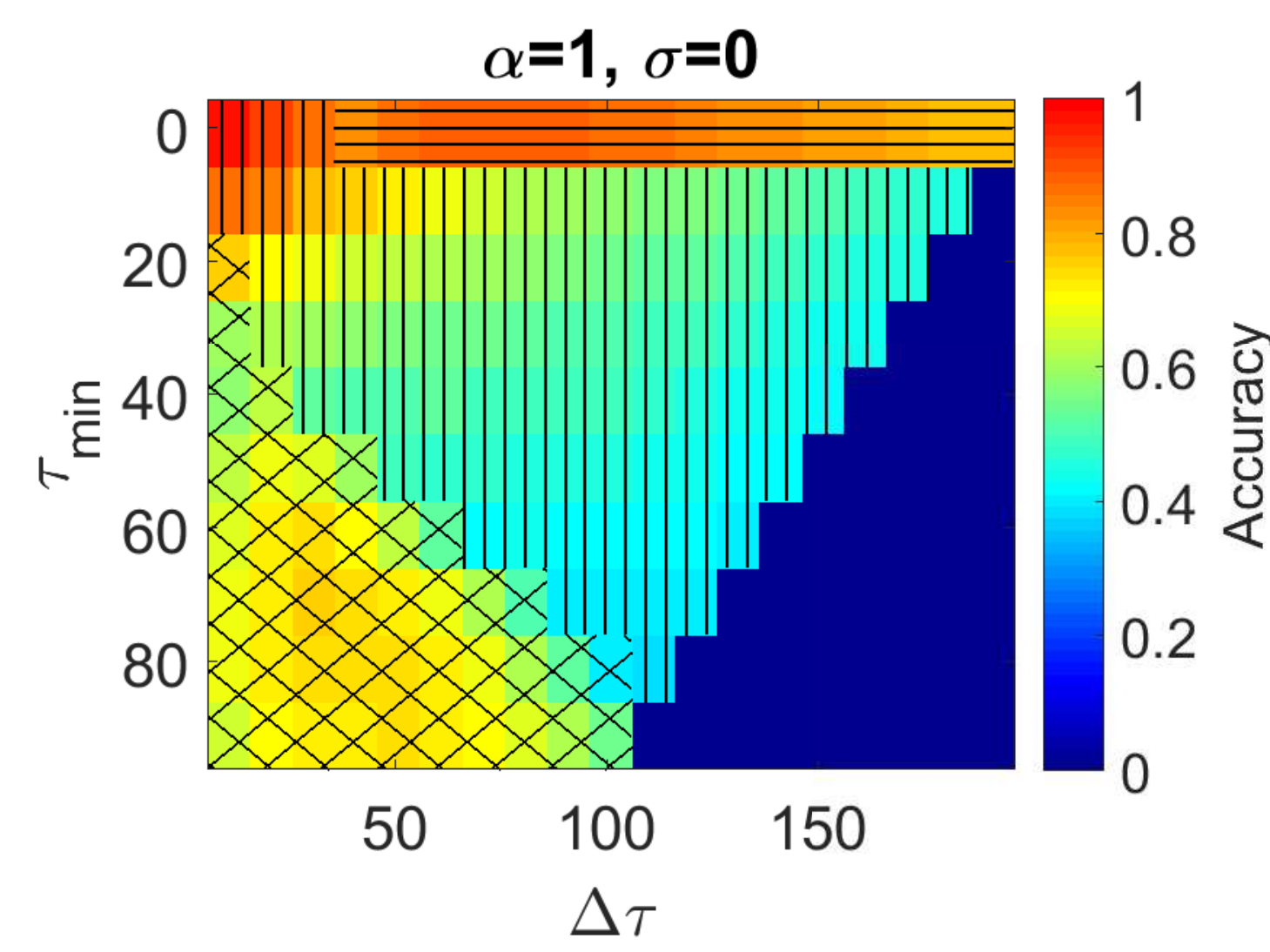}
\includegraphics[width=0.3\textwidth]{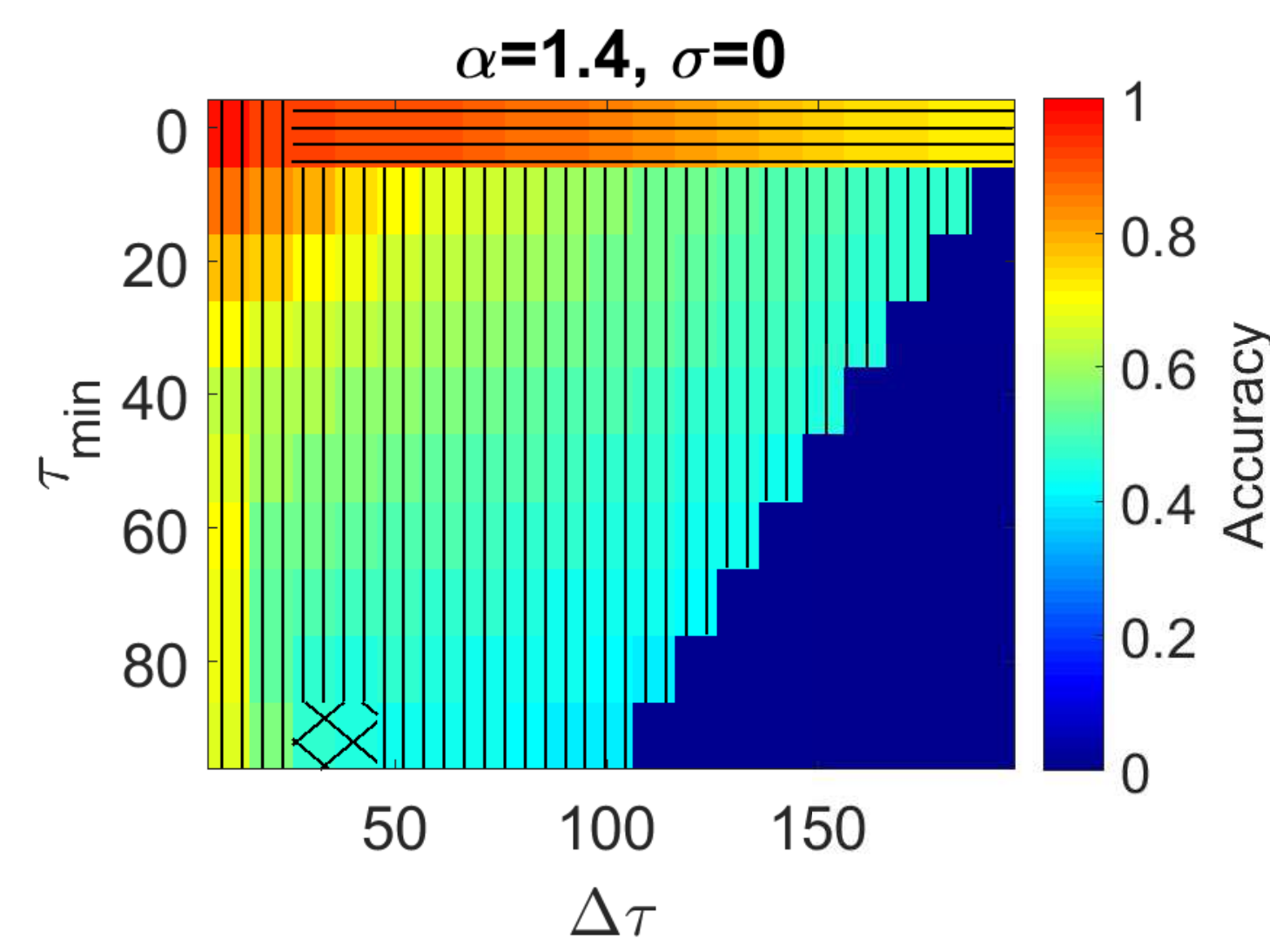}
\\
\includegraphics[width=0.3\textwidth]{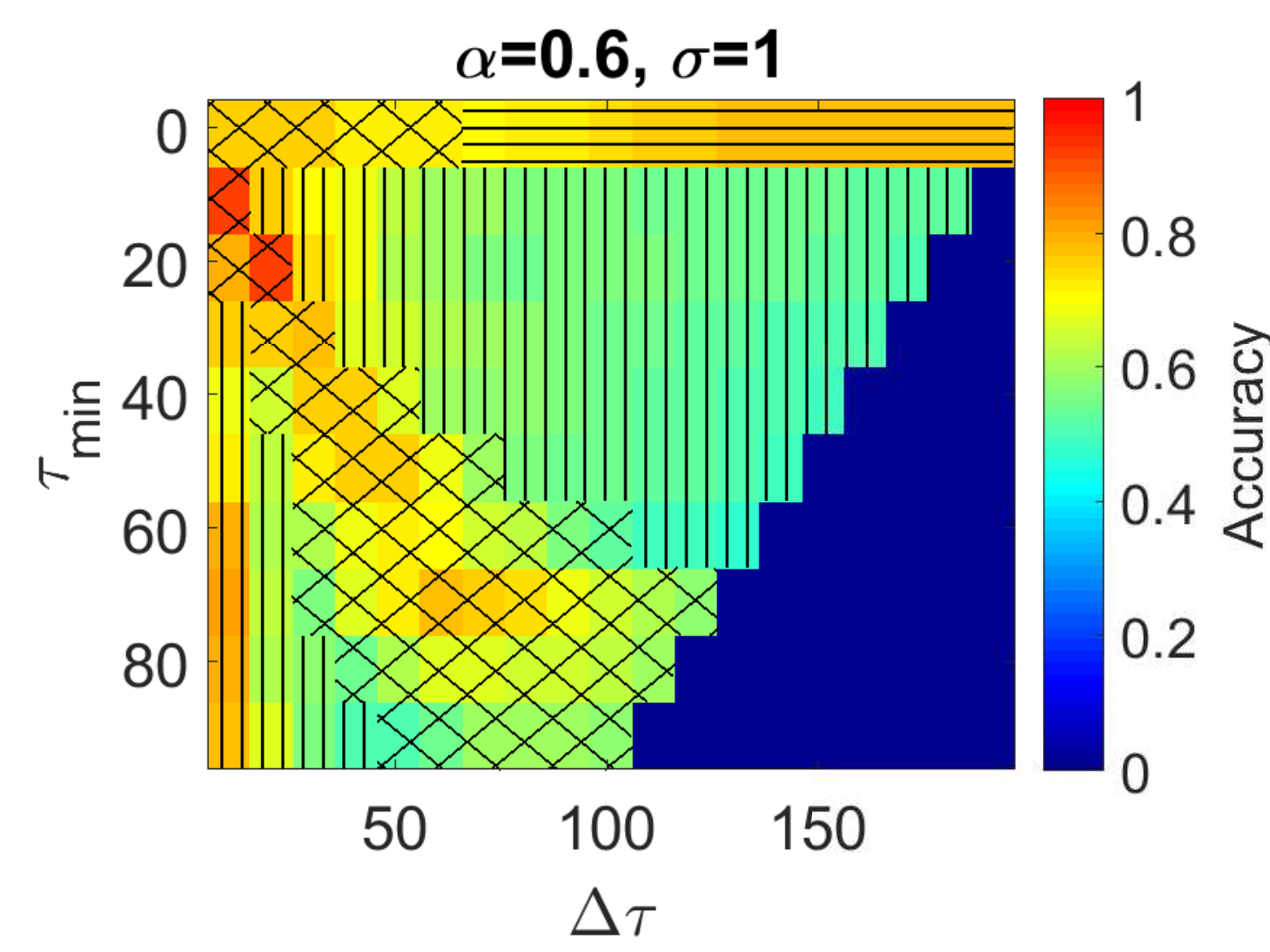}
\includegraphics[width=0.3\textwidth]{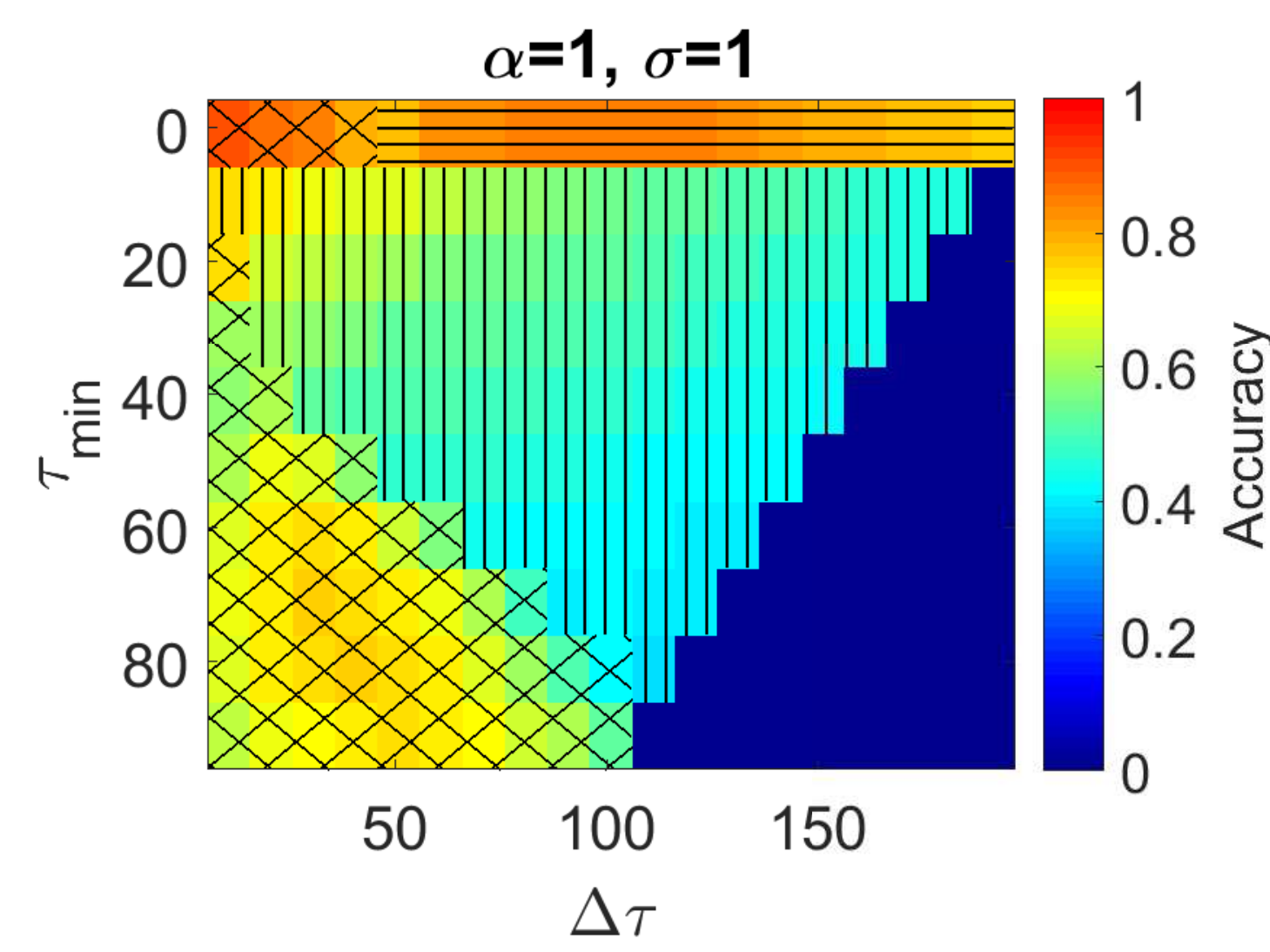}
\includegraphics[width=0.3\textwidth]{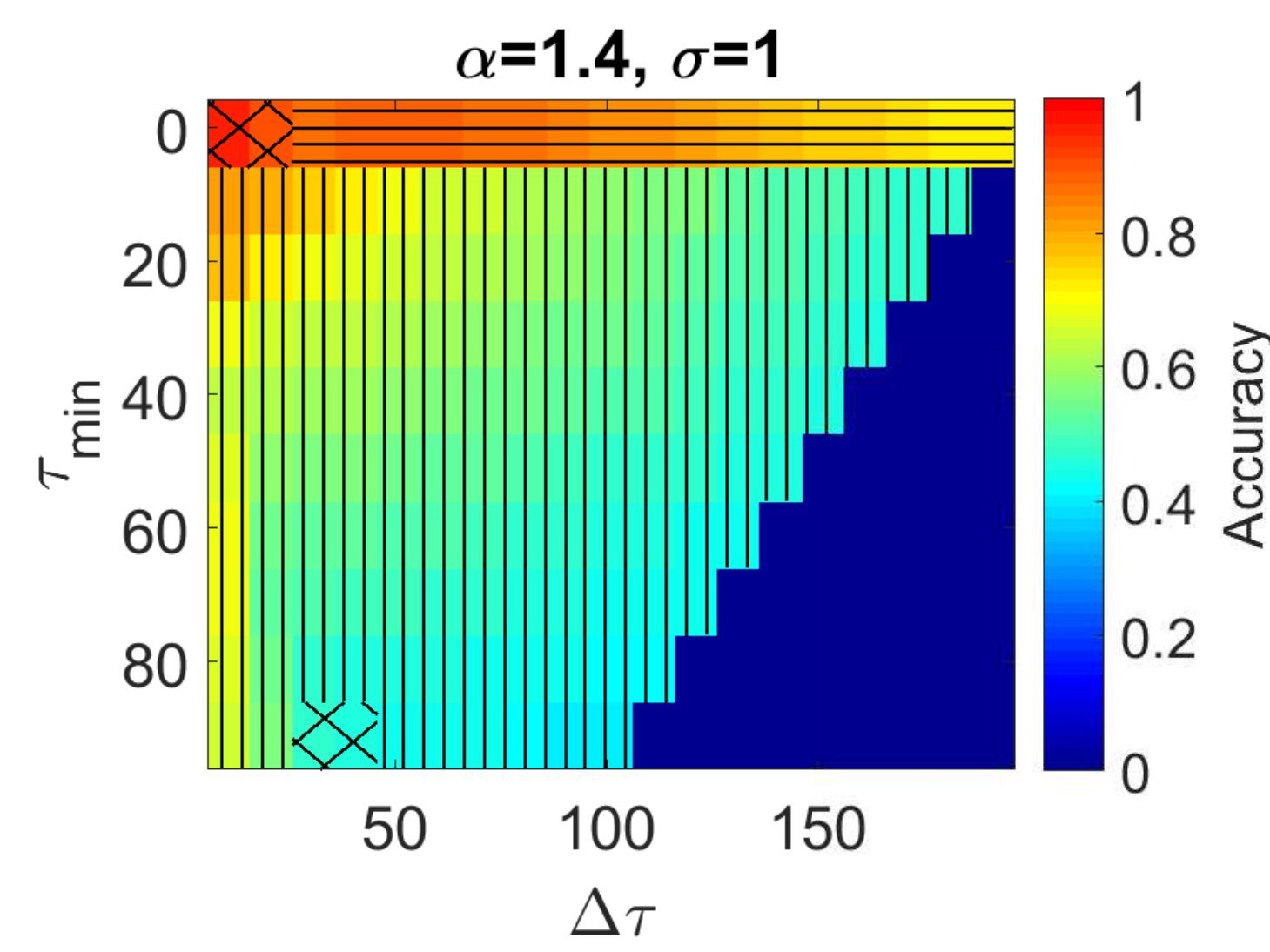}
\\
\includegraphics[width=0.3\textwidth]{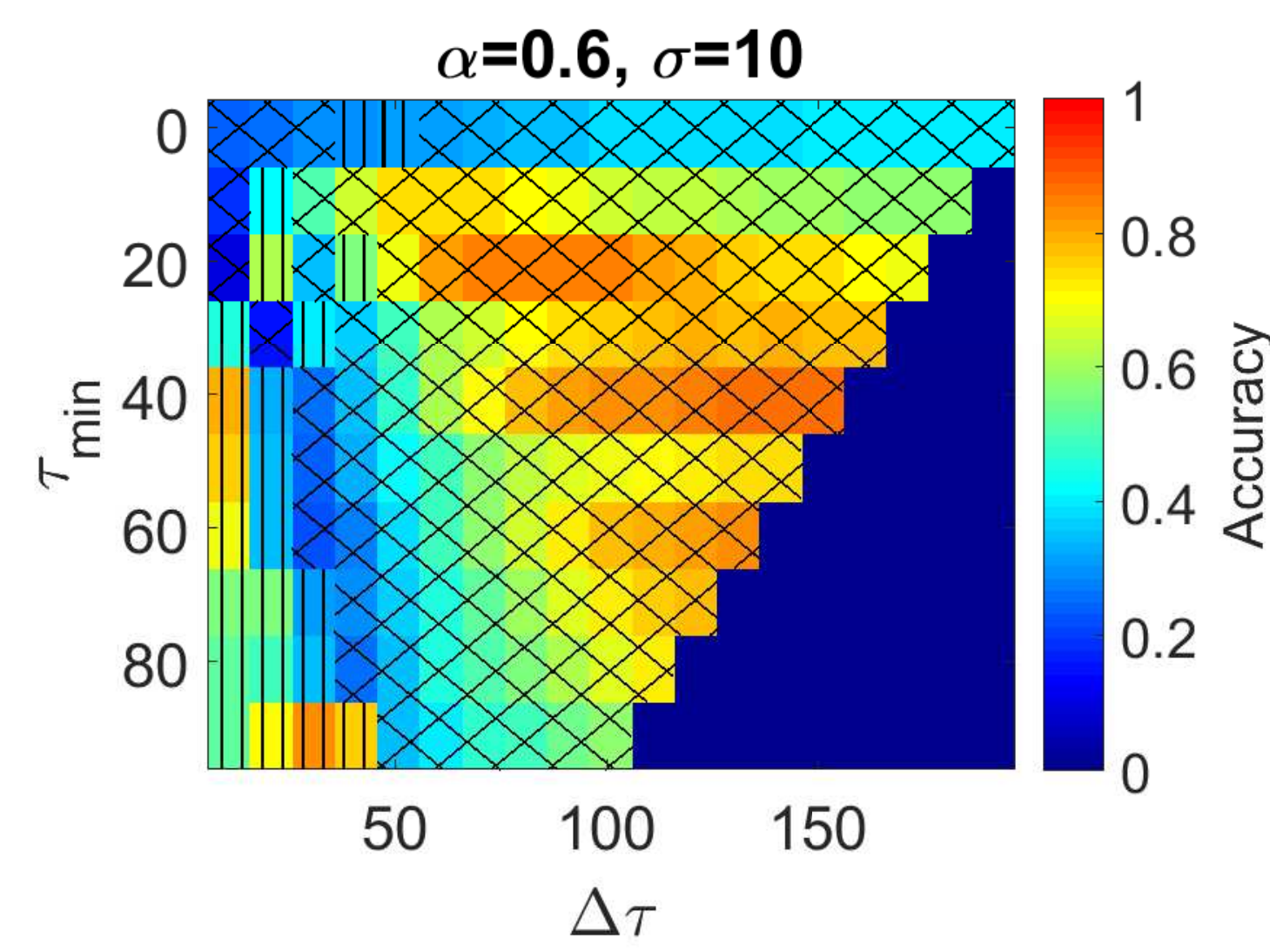}
\includegraphics[width=0.3\textwidth]{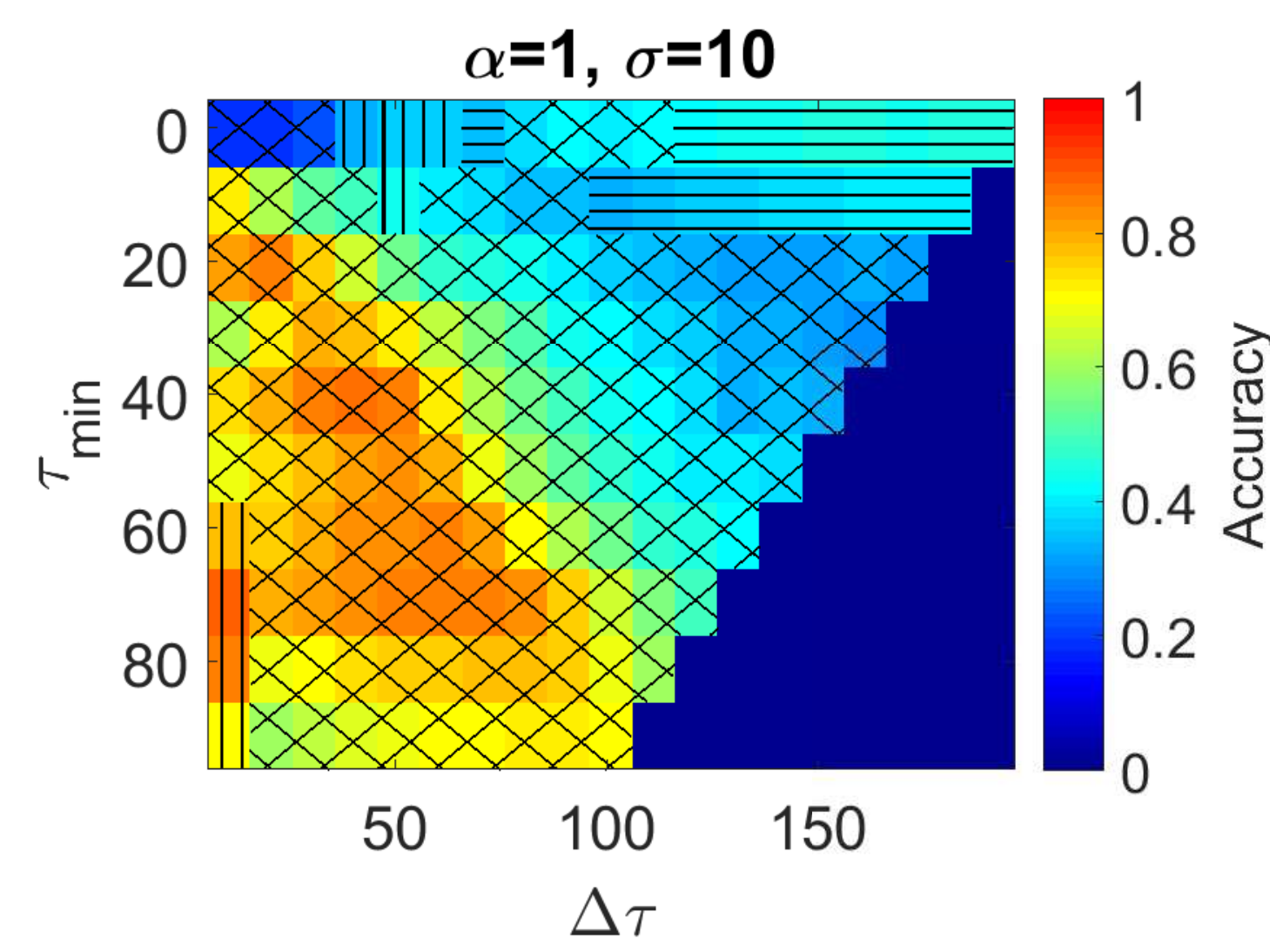}
\includegraphics[width=0.3\textwidth]{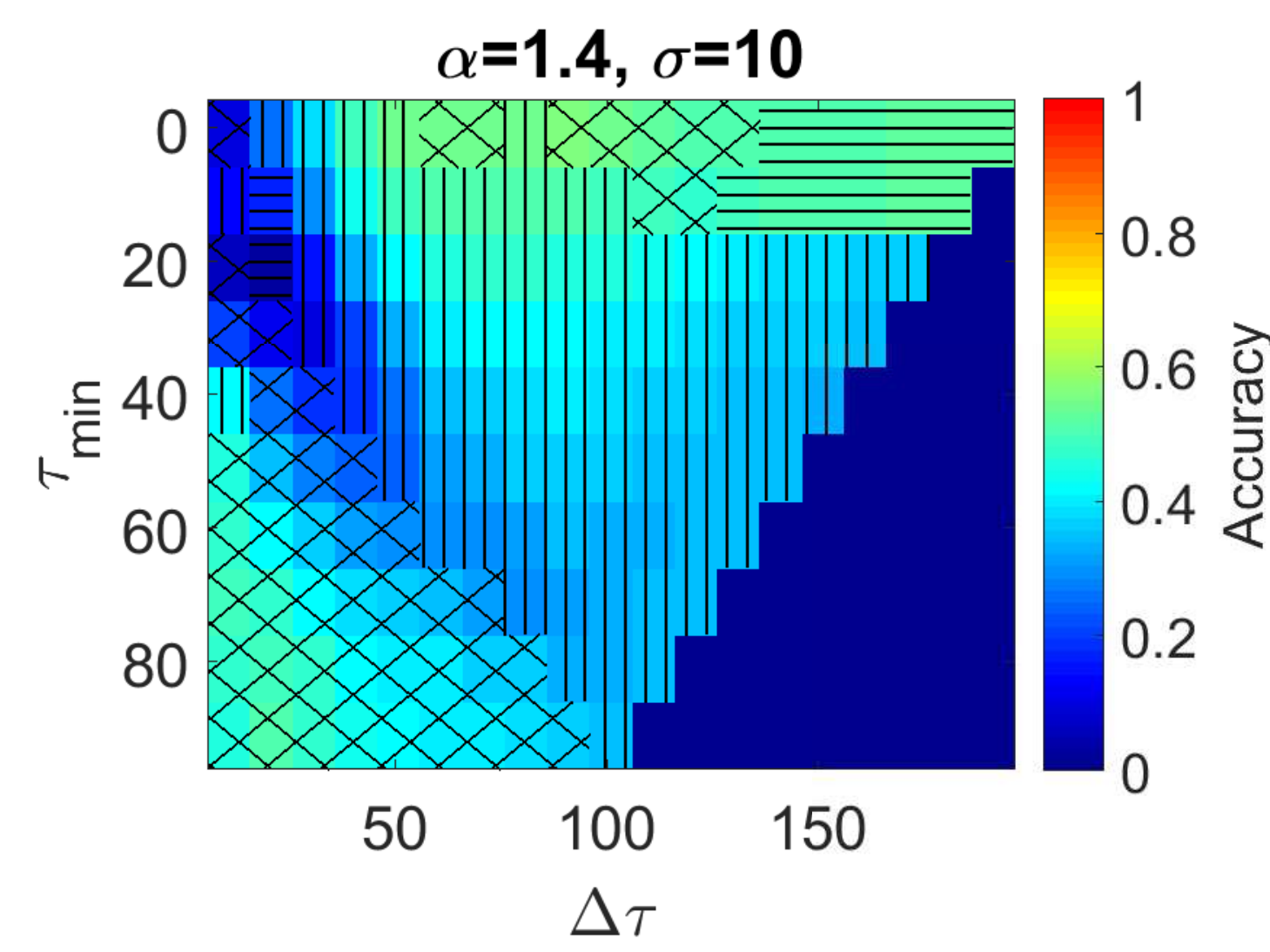}
\caption{
Percentage of estimations in the range
$\beta-0.2<\hat{\beta}<\beta+0.2$ for the best approach among the
three tested. From {\it left} to {\it right} $\beta =\lbrace
0.6,1,1.4\rbrace$, from {\it top} to {\it bottom} the standard
deviation of white noise in the range $\sigma=\lbrace 0,1,10\rbrace$
with $D_{\beta}=1/2$ and $N=1000$.  Each pair
$[\tau_{\min},\Delta\tau$], where $\Delta\tau=\tau_{\max}-\tau_{\min}$
with $\tau_{\min} \in[1,11,\ldots,191]$ and $\tau_{\max}
\in[11,\ldots,201]$, are considered. Dashed patterns highlight the
most accurate approach with {\it Horizontal, vertical } and {\it
crossed} lines respectively associated to Approaches I, II, and III;
colors highlight the corresponding best Accuracy score from dark blue
($Accuracy=0$) to light red ($Accuracy=100\%$); dark blue triangular
regions ($\tau_{\max}>20N/100$) were not tested.}
\label{Table_N1000}
\end{figure*}

First, we determine which approach from Section \ref{sec_2} is the
most accurate for each couple $\tau_{\min},\tau_{\max}$ and level of
noise $\sigma$.  For a better comparison, it is convenient to replace
$\tau_{\max}$ by the time window width
\begin{equation*}
\Delta \tau = \tau_{\max} - \tau_{\min} .
\end{equation*}
Figure \ref{Table_N1000} shows the result for long trajectories
($N=1000$). In the case without noise (first row), the classical
approach (Approach I) performs rather poorly as it can outperform
other approaches only in the case where $\tau_{\min}=1$ and $\Delta
\tau >110,40,30$ for $\beta=0.6,1,1.4$, respectively.  The Approach
III gives satisfactory results in the lower triangle where roughly
$\tau_{\min}>\Delta\tau$, from anti-persistent to diffusive motion
($\beta\leq 1$) while it is Approach II that is better for $\beta>1$
in this region. In all other situations, Approach II is the best.

Conclusions drawn from the first row of Figure \ref{Table_N1000} are
also applicable to small noise (second row, $\sigma=1$). This is
understandable as the TAMSD is affected by this level of noise only
around $\tau\approx 1$ (see Figure \ref{fig:plot_MSD_plus_noise}).  In
contrast, for large noise ($\sigma=10$, third row), the TAMSD is
affected by the noise on a longer time range, and the results are
different: (i) for $\beta=0.6$ the best estimation is achieved by
increasing both $\tau_{\min}$ and $\Delta \tau$; (ii) for $\beta=1$
the best score is achieved by increasing $\tau_{\min}$ but keeping
$\Delta \tau$ not too high; (iii) for $\beta=1.4$, the quality of the
estimations is poor in every case.  The last result is
counter-intuitive as the impact of noise is reduced as $\beta$
increases (see Eq. (\ref{MSD_small_noise})) but this reduction is not
enough at small $\tau$, for instance, noise still presents one third
of the MSD ($2D\tau_c^{1.4}/\sigma^2=2$) at $\tau_c\approx 44$.  At
longer lag-time $\tau$ (but still with $\tau <N/20$), the positive
auto-correlations slow down the self-averaging so the distribution of
the TAMSD is wider \cite{testexponent}. The combination of noise and
the wider distribution of TAMSD prevents obtaining a correct
estimation of the exponent with trajectories of length $N = 1000$.

\begin{figure*}
\includegraphics[width=0.3\textwidth]{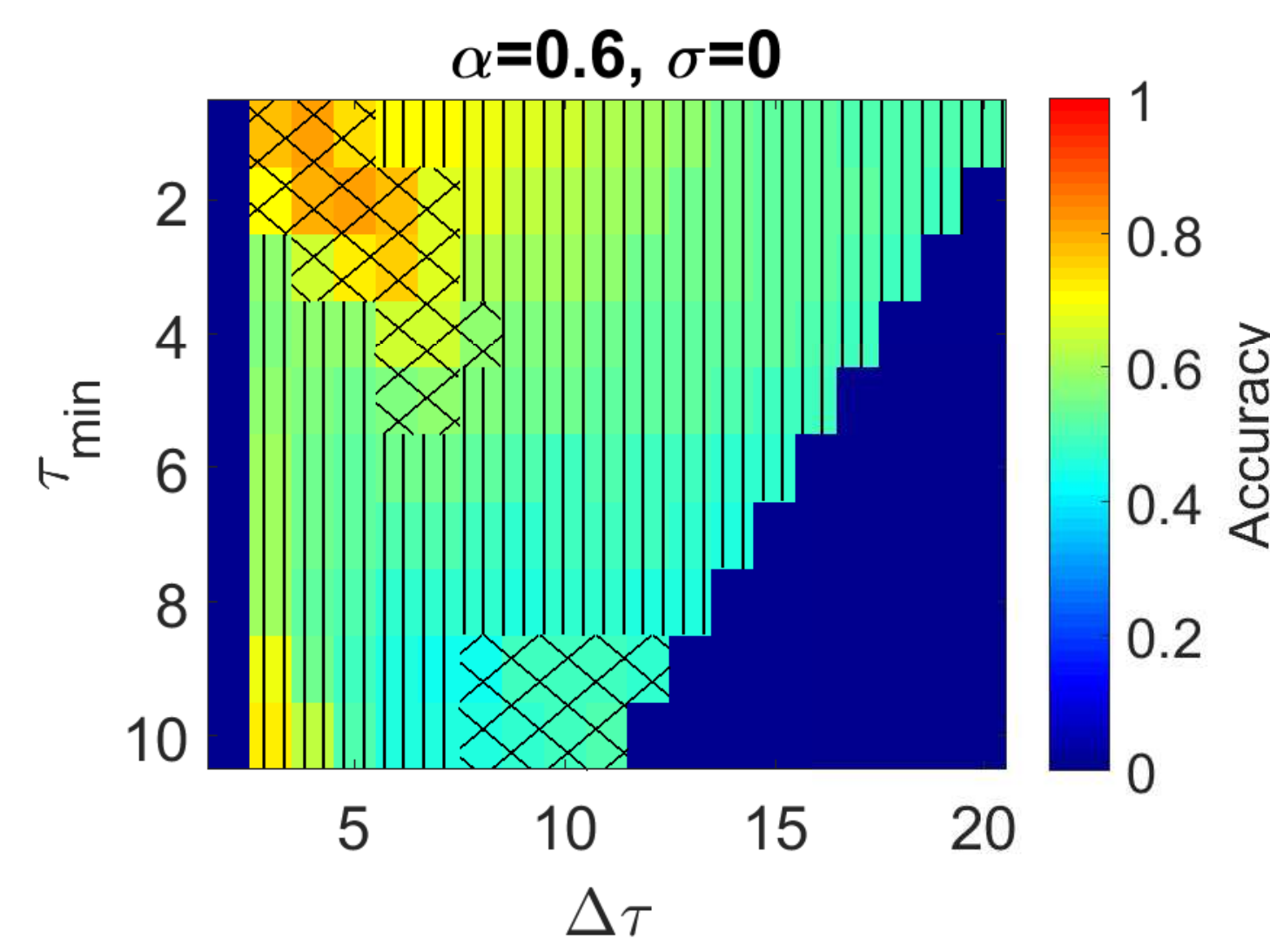}
\includegraphics[width=0.3\textwidth]{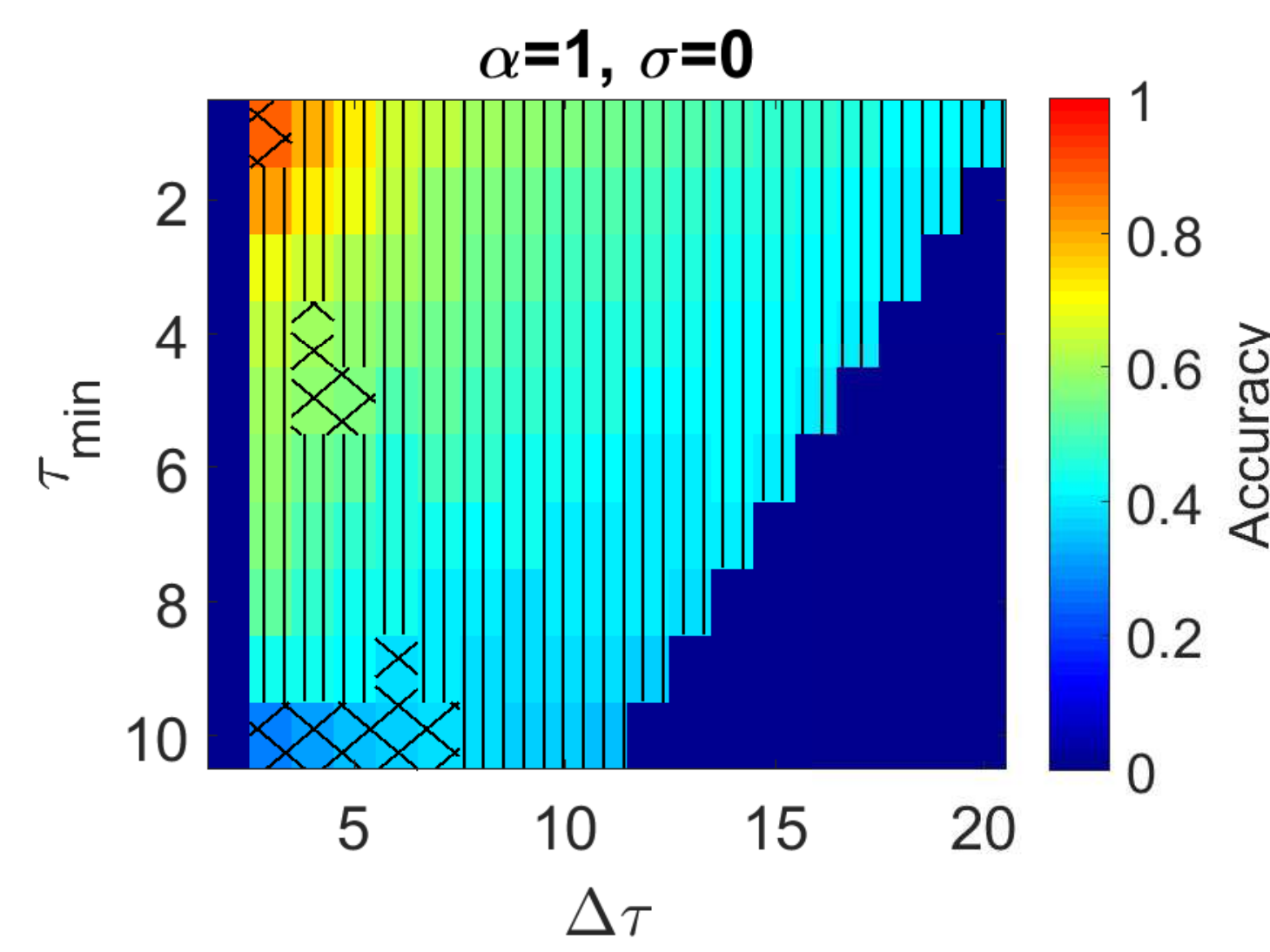}
\includegraphics[width=0.3\textwidth]{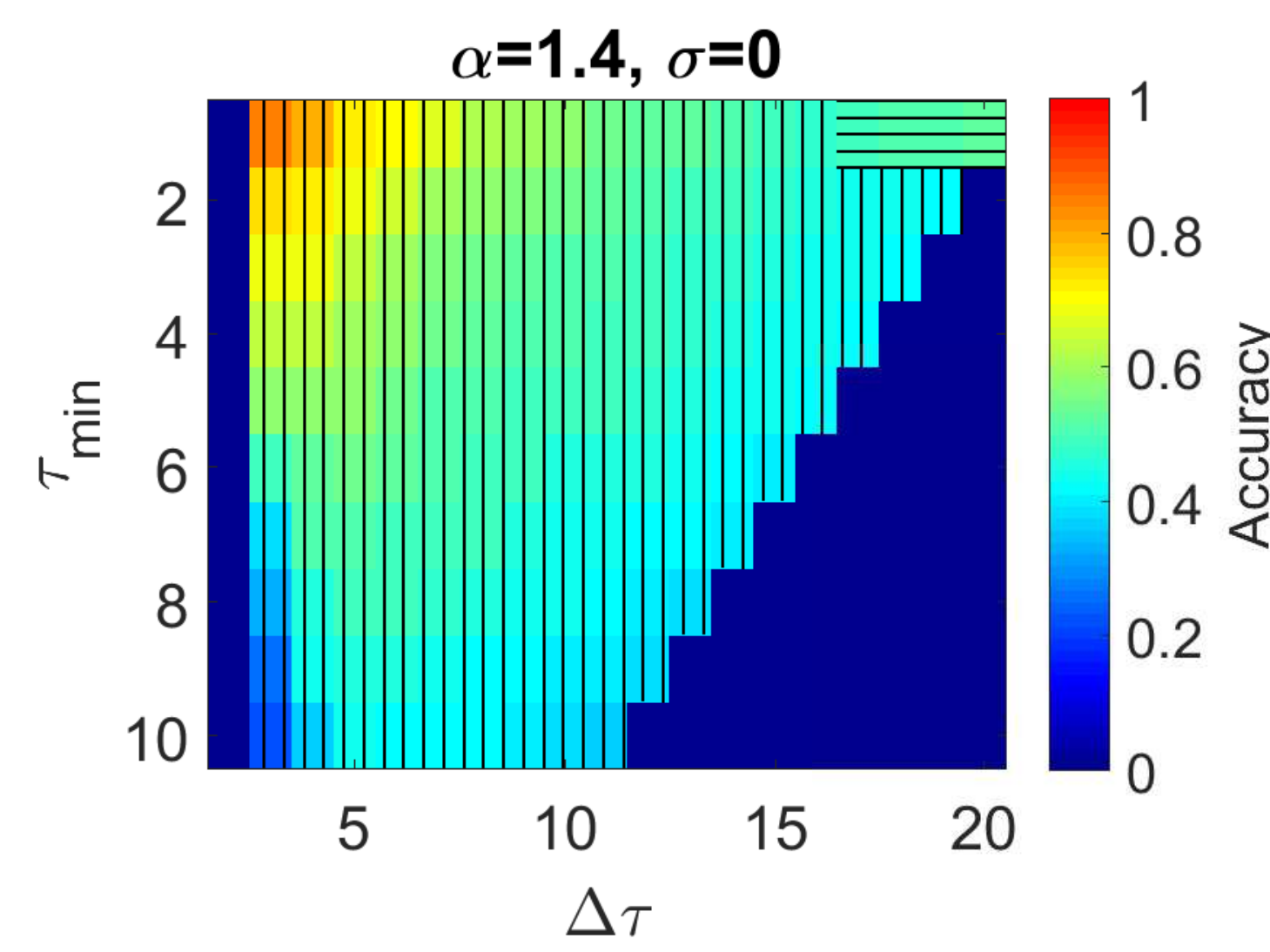}
\\
\includegraphics[width=0.3\textwidth]{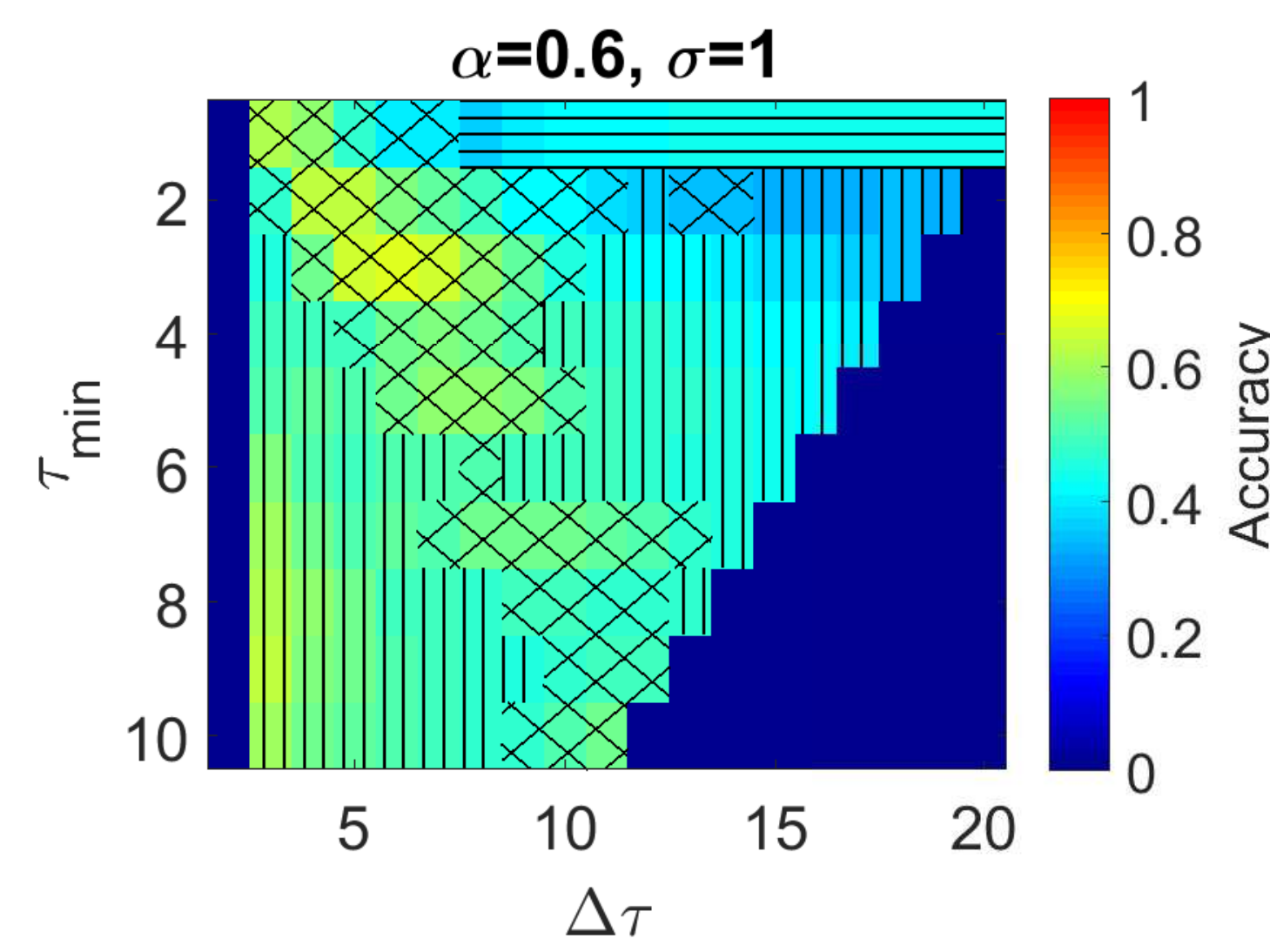}
\includegraphics[width=0.3\textwidth]{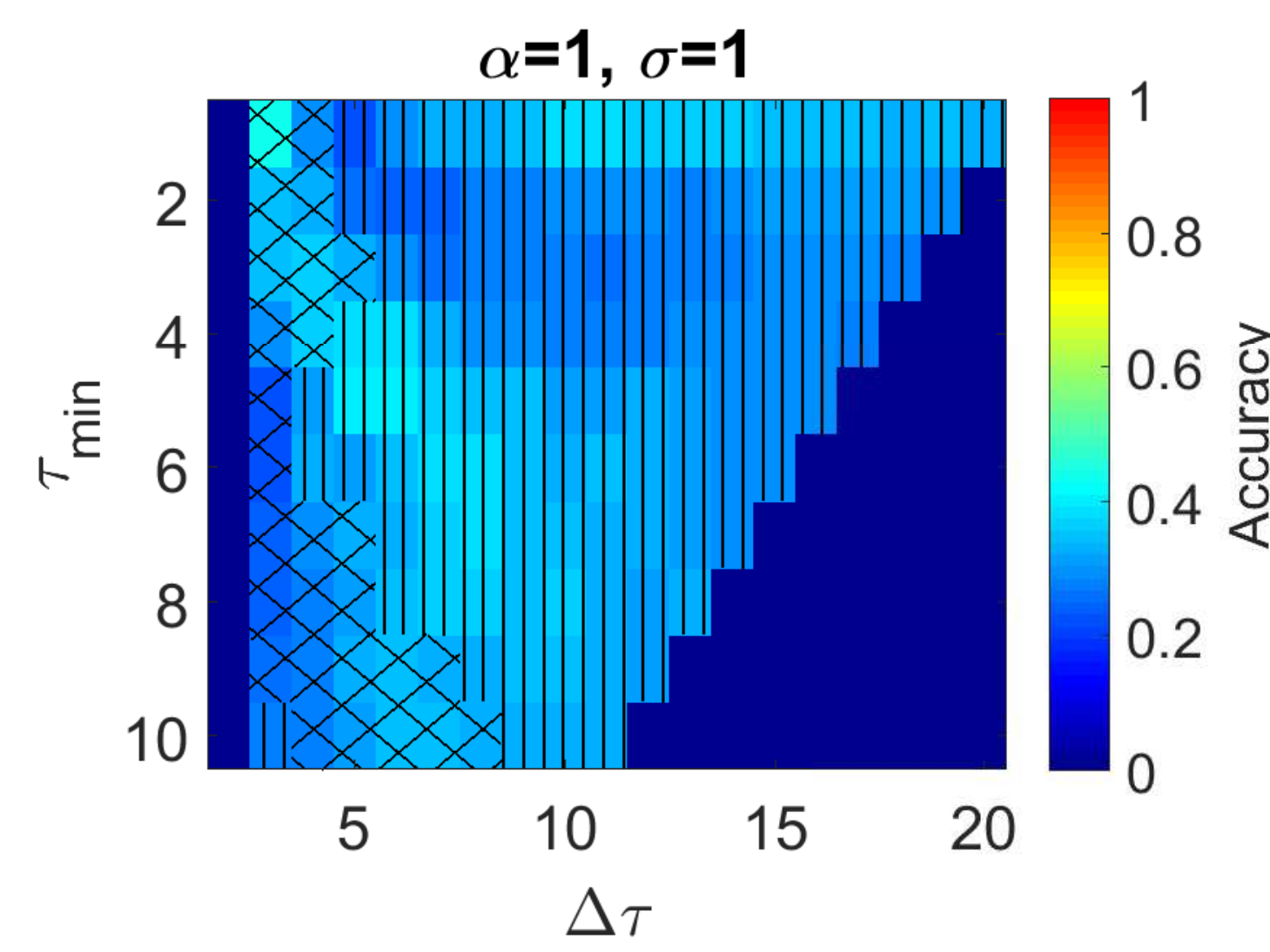}
\includegraphics[width=0.3\textwidth]{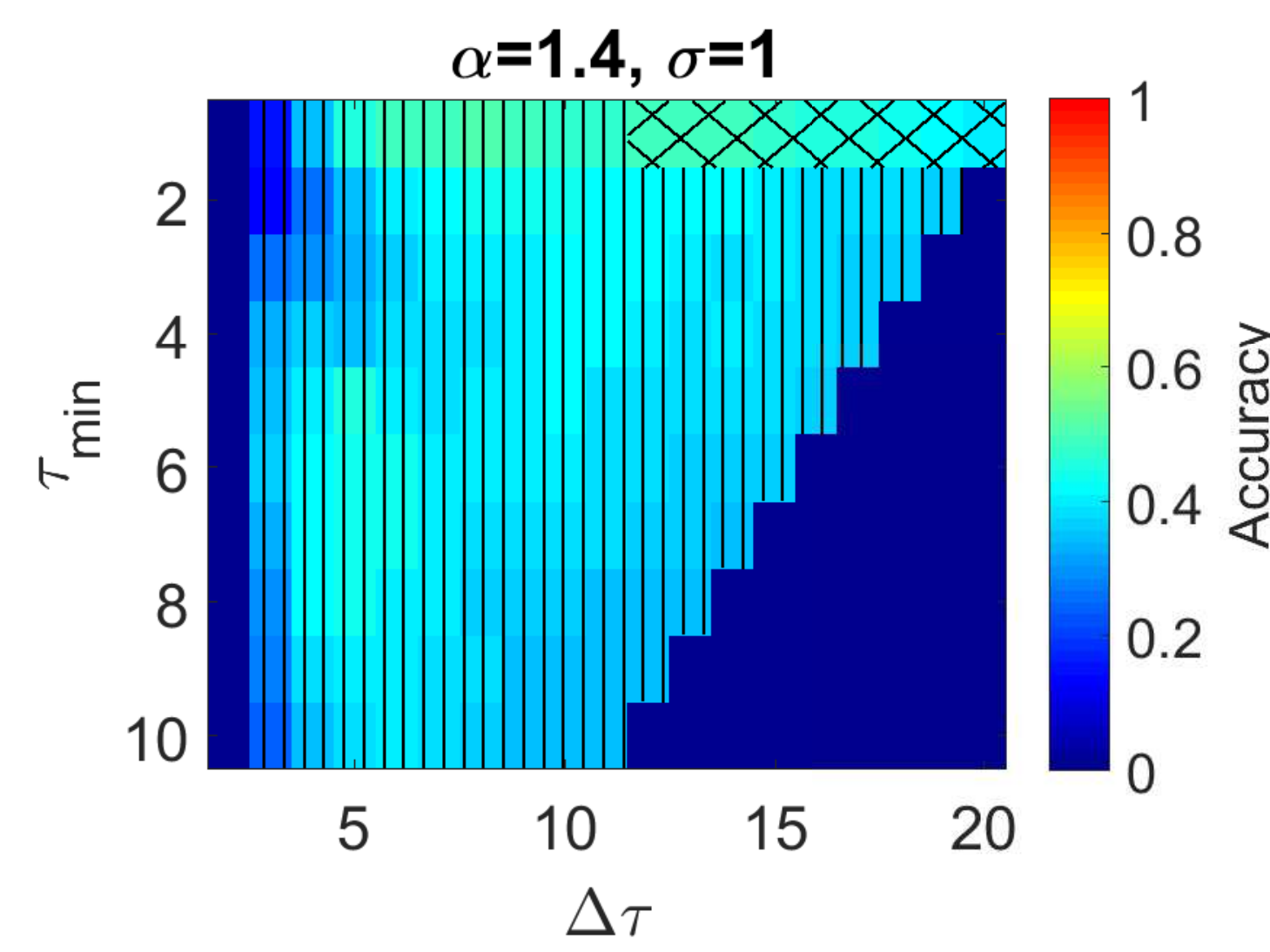}
\\
\includegraphics[width=0.3\textwidth]{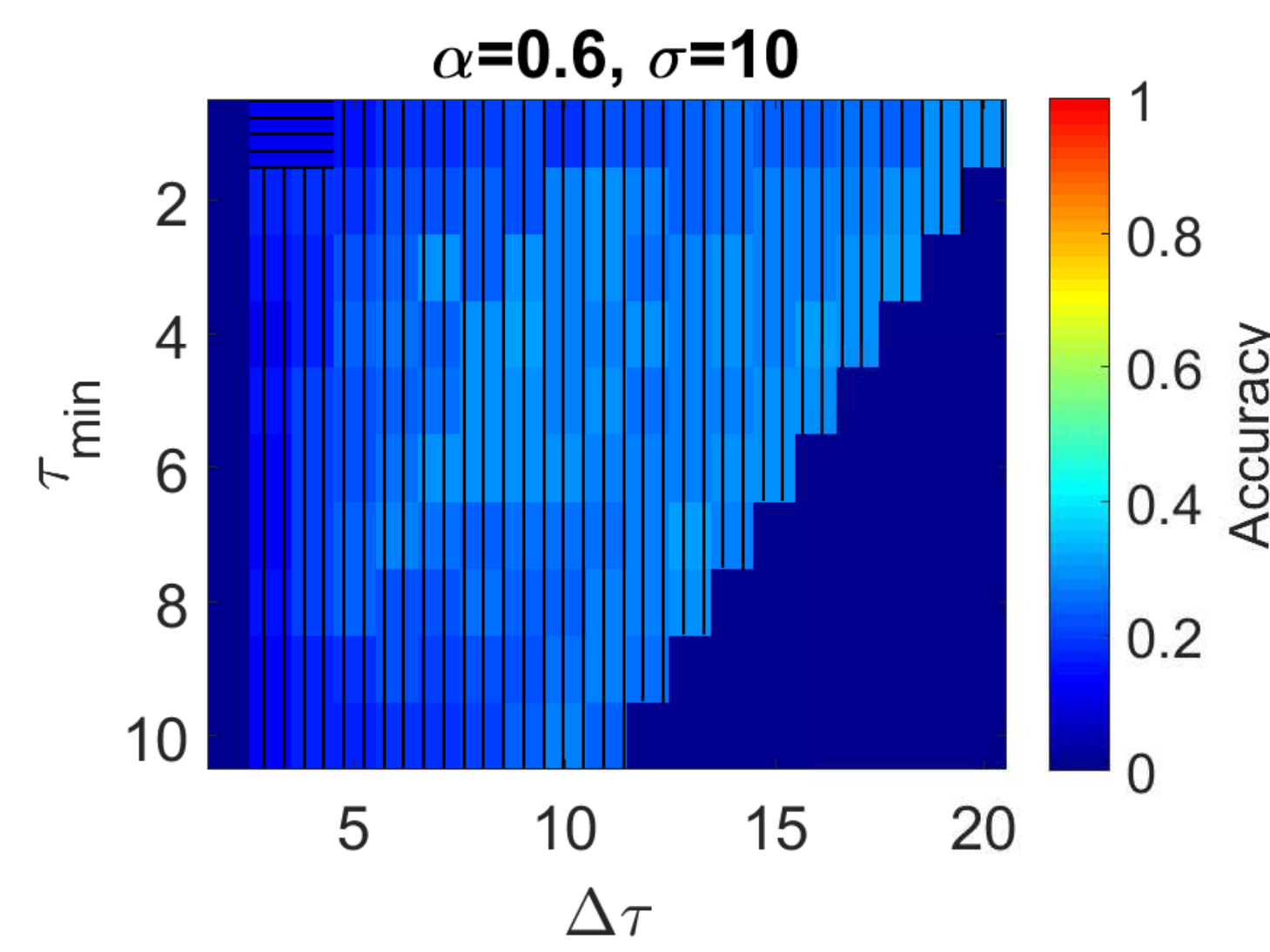}
\includegraphics[width=0.3\textwidth]{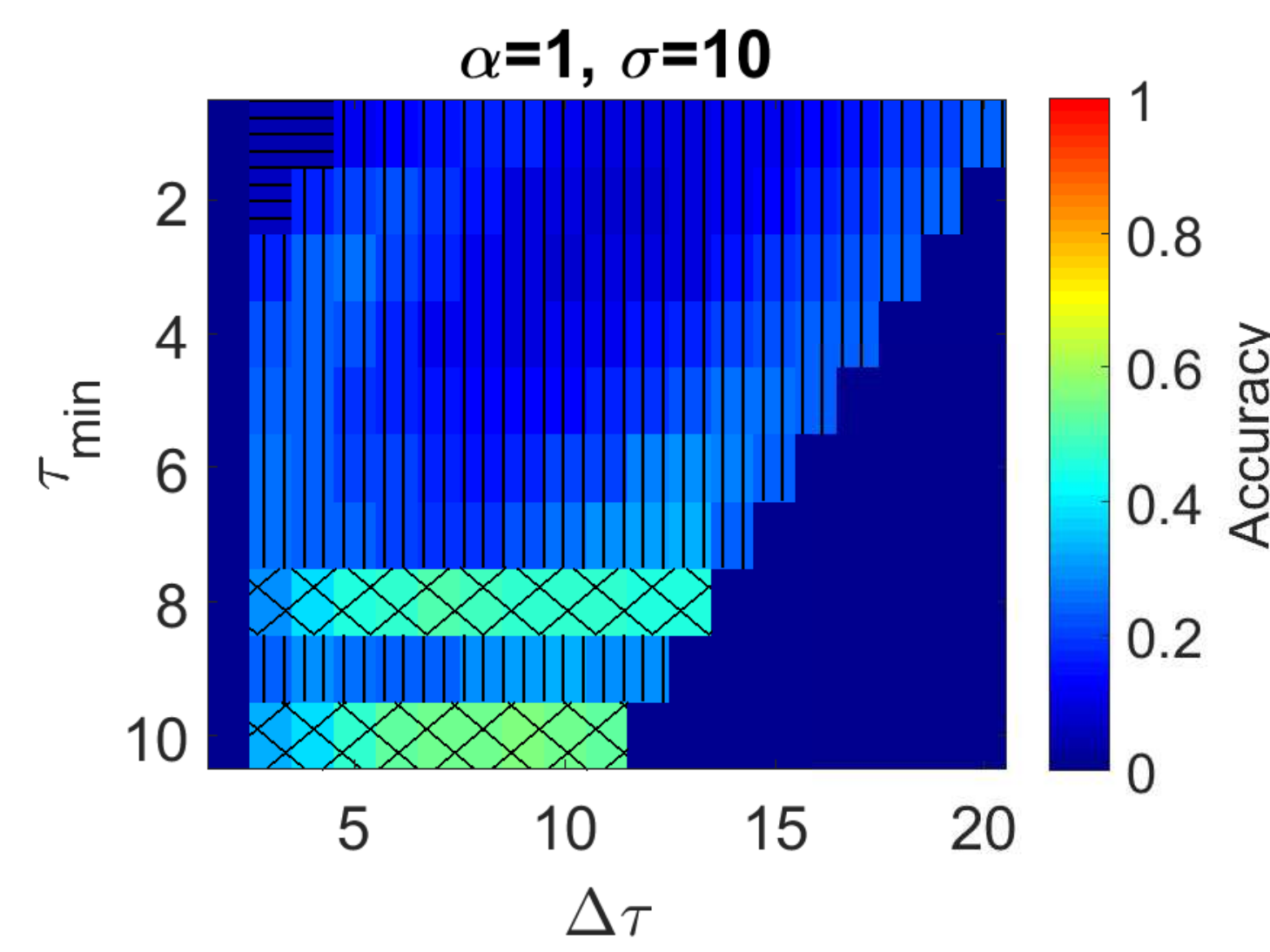}
\includegraphics[width=0.3\textwidth]{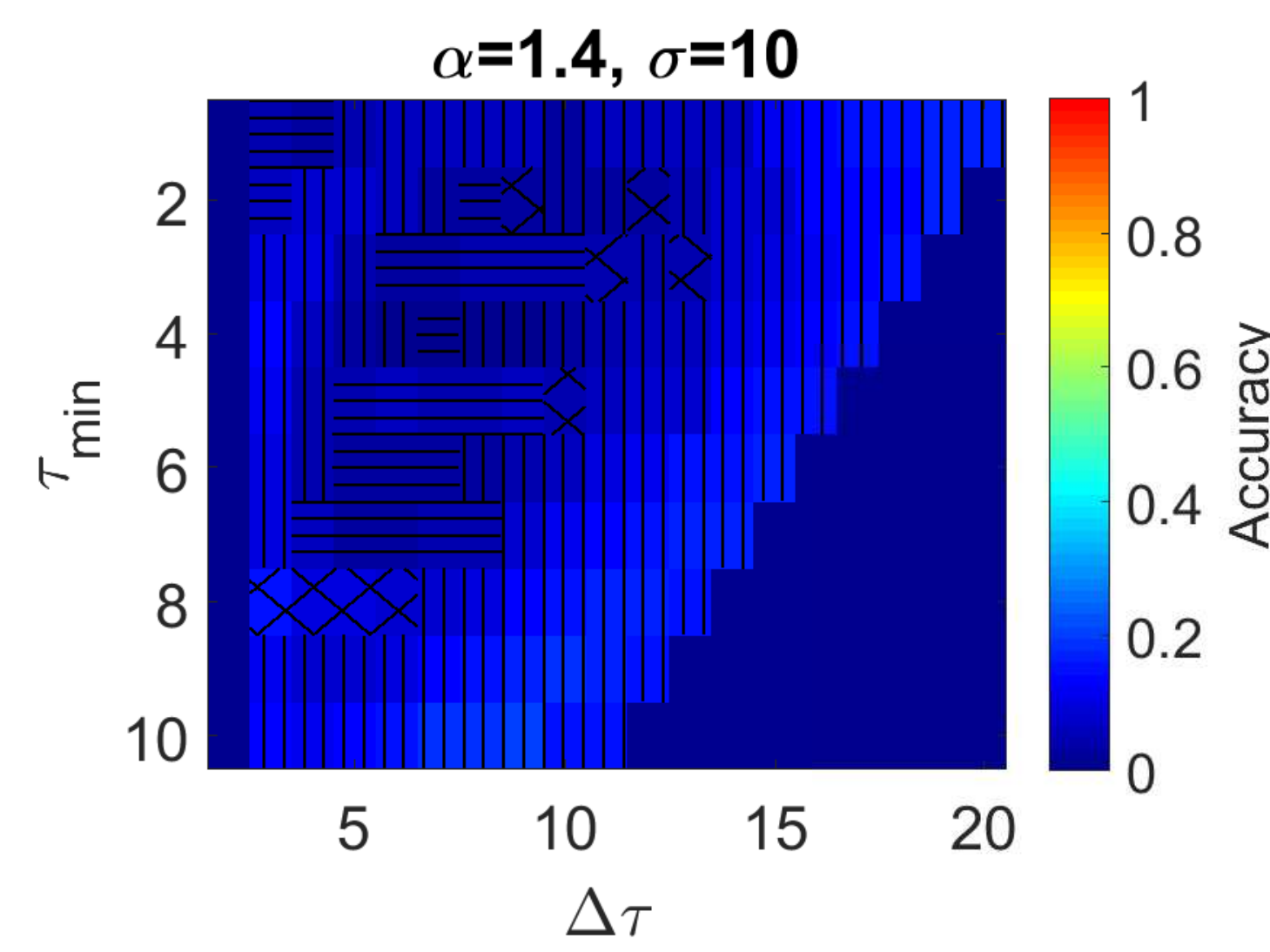}
\caption{
Percentage of estimations in the range
$\beta-0.2<\hat{\beta}<\beta+0.2$ (Accuracy) for the best approach
among the three tested. From {\it left} to {\it right} $\beta =
\lbrace 0.6,1,1.4\rbrace$, from {\it top} to {\it bottom} the standard
deviation of white noise in the range $\sigma=\lbrace 0,1,10\rbrace$
with $D_{\beta}=1/2$ and $N=100$.  Each pair
$[\tau_{\min},\Delta\tau$], where $\Delta\tau=\tau_{\max}-\tau_{\min}$
with $\tau_{\min} \in[1,2,\ldots,19]$ and $\tau_{\max}
\in[3,\ldots,20]$, are considered. Dashed patterns highlight the most
accurate approach with {\it Horizontal, vertical } and {\it crossed}
lines respectively associated to Approaches I, II, and III; colors
highlight the corresponding best Accuracy score from dark blue
($Accuracy=0$) to light red ($Accuracy=100\%$); dark blue triangular
regions ($\tau_{\max}>20N/100+1$) were not tested.  }
\label{Table_N100}
\end{figure*}

Figure \ref{Table_N100} shows the results in the same conditions for
trajectories of length $N=100$.  Such short trajectories are often
encountered in biological applications.  In this case, the
distribution of the TAMSD is wider, thus the estimation is more
difficult. When there is no noise, it is still possible to achieve a
good estimation for small $\tau_{\min}$ and $\Delta \tau$ while the
presence of even mild noise makes the estimation unreliable. In the
regime of strong noise, the estimation is so bad that it would make no
difference to uniformly pick an exponent in the range $\beta\in[0,2]$.
Thus, for short trajectories the TAMSD is not appropriate.

What are the best approach and the optimal estimation parameters that
maximize the accuracy?  Looking at Figs. \ref{Table_N1000} and
\ref{Table_N100}, one can see that there is neither ``the best
approach'', nor the unique optimal values for $\tau_{\min}$ and
$\tau_{\max}$.  In turn, we can determine the best approach and the
optimal parameters for each combination of the trajectory length $N$,
exponent $\beta$ and level of noise $\sigma$.  The results are
gathered in Table \ref{tab:Best_accuracy}.  Strikingly, the commonly
used Approach I is nowhere the best. For $N=1000$, when there is no
noise, the best choice is Approach II in every case, with
$\tau_{\min}=1$ and $\Delta\tau=10$ (note that the optimal value for
$\Delta\tau$ can be even smaller, due to the discrete exploration of
the parameters space).  In the presence of noise, the Approach III is
the best, with progressively increasing $\tau_{\min}$ and $\Delta
\tau$ as the noise level increases.  For $N=100$, the noise impacts
significantly the accuracy. Even for $\sigma=1$, accuracy drop to
$\approx 50\%$ emphasizing that precise estimation based on such a
short trajectory requires a very good experimental signal to noise
ratio.

\begin{table*}
  \begin{center}

   \begin{tabular}{| c || c || c | c | c | c || c | c | c | c || c | c | c | c | }
\hline
\multirow{2}*{$N$} &   \multirow{2}*{$\sigma$} & \multicolumn{4}{c||}{$\beta=0.6$} & \multicolumn{4}{c||}{$\beta=1$} & \multicolumn{4}{c|}{$\beta=1.4$}
\\   
 \cline{3-14}
 & & Approach & $\tau_{\min}$ & $\Delta\tau$ & $A(\%)$ & Approach & $\tau_{\min}$ & $\Delta\tau$ & $A(\%)$ & App. & $\tau_{\min}$ & $\Delta\tau$ & $A(\%)$ \\ 
  \cline{1-14}
   \cline{1-14}
\multirow{3}*{$1000$}  &       $0$ & II & 1 & 10 & 98 & II & 1 & 10 & 97 & II & 1 & 10 & 98 \\  \cline{2-14}
   &  $1$ & III & 11 & 10 & 92 & III & 1 & 10 & 89 & III & 1 & 10 & 86 \\
      \cline{2-14}
    & $10$ & III & 41 & 150 & 86 & III  & 71 & 10 & 88 & III & 1 & 90 &	55 \\

  \cline{1-14}
   \cline{1-14}
 \multirow{3}*{$100$} &     $0$ & III & 2 & 4 & 82 & III & 1 & 2 & 88 & II & 1 & 2 & 86 \\ \cline{2-14}
 &    $1$ & III & 3 & 5 & 66 & III & 1 & 2 & 44 & III & 1 & 7 & 50 \\
    \cline{2-14}
&     $10$ & II & 7 & 12 & 32 & III & 10 & 8 & 56 & II & 10 & 8 & 21 \\
     \hline
   \end{tabular}
 \end{center}
\caption{
Summary of the best approaches for two lengths of trajectory, $N=100$
and $N=1000$, with three level of noise $\sigma=0,1,10$, for three
cases of the exponent $\beta=0.6,1,1.4$. The table shows the best
approach with the corresponding parameters $\tau_{\min}$ and
$\Delta\tau=\tau_{\max}-\tau_{\min}$ with
$\tau_{\min}\in[1,N/100+1,\ldots,19N/100+1]$ and
$\tau_{\max}\in[N/100+1,\ldots,20N/100+1]$, and the corresponding
accuracy $A(\%)$. }
\label{tab:Best_accuracy}
 \end{table*}

\section{Conclusions}

We studied the problem of estimation of the anomalous diffusion
exponent for processes in which a ``pure'' anomalous diffusion model
is corrupted by independent noise.  We propose two alternative
approaches that can be used for estimating anomalous diffusion
exponent. We indicate their advantages and limitations and check their
efficiency by Monte Carlo simulations. We show that the classical
estimation fails in every case.  Moreover, none of the approaches is
the best for all cases. We indicate how the model parameters, as well
as parameters of the estimation techniques, may influence the
results. The presented discussion and results can be useful for a more
reliable statistical analysis of single-particle trajectories in cell
biology and other fields.

\section*{Acknowledgments}
\noindent 
D. S. Grebenkov acknowledges the support under Grant
No. ANR-13-JSV5-0006-01 of the French National Research Agency. \\
A. Grzesiek and A. Wy{\l}oma{\'n}ska would like to acknowledge a
support of NCN OPUS Grant No.\\ UMO-2016/21/B/ST1/00929 ``Anomalous
diffusion processes and their applications in real data modeling".
\appendix

\section{Properties of the $\hat{\beta}$  in Approach I}
The expected value of the estimator $\hat{\beta}$ defined in \eqref{est_2} takes the form
\begin{widetext}
\begin{eqnarray} \nonumber
	\E\{\hat{\beta}\}&=&\frac{n\sum_{\tau=\tau_{\min}}^{\tau_{\max}}{\ln(\tau)\ln(E\{M_N(\tau))\}}-\sum_{\tau=\tau_{\min}}^{\tau_{\max}}{\ln(\tau)}\sum_{\tau=\tau_{\min}}^{\tau_{\max}}{E\{\ln(M_N(\tau))\}}}{n\sum_{\tau=\tau_{\min}}^{\tau_{\max}}{\ln^2(\tau)}-\left(\sum_{\tau=\tau_{\min}}^{\tau_{\max}}{\ln(\tau)}\right)^2}
\end{eqnarray}
\end{widetext}
which for the infinitely long trajectory gives
\begin{equation}
\E\{\hat{\beta}\} \stackrel{N\rightarrow \infty}{\rightarrow} \beta.
\end{equation}
One can also calculate the exact expression of $\Var\{\hat{\beta}\}$.
As the idea is similar to \cite{testexponent} we do not repeat long
calculations here.  From the Cauchy-Schwarz inequality and the fact
that $\Var\{\ln(M_N(\tau))\}\stackrel{N\rightarrow
\infty}{\rightarrow}0$, we get that the variance of the estimator
vanishes at long time
\begin{equation}
\Var\{\hat{\beta}\} \stackrel{N\rightarrow \infty}{\rightarrow}0.
\end{equation}
This makes the estimator consistent and asymptotically unbiased in the
case without noise.  The conclusions hold in the presence of noise in
the region where diffusion is dominant for
$\tau_{\min}\gg(\sigma^2/D_{\beta})^{1/\beta}$.  However, one has
never access to infinitely long trajectories in real conditions, so
the estimator has an intrinsic distribution.

\section{Non-linear fitting}\label{sec:NL_fit}

Non-linear fitting in Approaches II and III consists in finding the
parameters $\hat{D}_{\beta}$, $\hat{\beta}$ and possibly
$\hat{\sigma}$ (see (\ref{f1}) and (\ref{f2})) that minimize the sum
of squared errors. For a non-linear problem, there is no explicit
expression for the estimator, and one has to perform the minimization
procedure by numerical methods. In this article, we use a trust region
method \cite{Coleman1994,Coleman1996} to perform non-linear least
square fitting with Matlab. In order to reduce the calculation time
and avoid nonphysical values of parameters, some constraints are
imposed on the parameters. All parameters are positive, the exponent
$\hat{\beta}$ cannot exceed the ballistic regime, $\hat{\beta} \leq
2$, and the noise is necessarily smaller than the TAMSD at $\tau=1$ so
$\hat{\sigma}^2\in [0,M_N(1)]$. There is no evident upper bound for
the generalized diffusion coefficient so we assume $\hat{D}_\beta\in
[0,\infty)$.  For the minimization procedure, a crucial point is the
choice of the stopping criterion $\epsilon$. The iteration is
interrupted when the relative change in the error function
$\frac{\vert\Upsilon_{i+1}-\Upsilon_i\vert}{1+\vert\Upsilon_i\vert}
<\epsilon$. Choosing $\epsilon$ too large forces the algorithm to stop
before convergence, resulting in poor estimation. Conversely, taking
$\epsilon$ too small makes the minimization longer because the random
nature of the TAMSD imposes a lower limit on the possible precision
obtained.  In our case, $M_N$ does not follow exactly the theoretical
MSD as the TAMSD, evaluated over a single realization of a stochastic
process of finite length $N$, is itself random. Thus one cannot expect
a perfect match between $\hat{M}$ and $M_N$, in other words, there is
a distribution of the minimum for the function $\Upsilon$ which is
determined by the fluctuations of the TAMSD, depending on
$\tau_{\min},\tau_{\max},N,$ and $\beta$, moreover the presence of a
white noise increases uncertainty and thus increases the optimal
$\epsilon$.  The best $\epsilon$ is the largest possible value for
which the estimation remains unchanged. In this article we chose
$\epsilon=0.01$ as a good compromise between speed and precision.

\end{document}